\documentclass{article} 
\usepackage[xcdraw, table]{xcolor}
\usepackage{iclr2022_conference,times}

\usepackage{amsmath,amsfonts,bm}

\def\eqref#1{equation~\ref{#1}}

\def\1{\bm{1}}

\DeclareMathAlphabet{\mathsfit}{\encodingdefault}{\sfdefault}{m}{sl}
\SetMathAlphabet{\mathsfit}{bold}{\encodingdefault}{\sfdefault}{bx}{n}

\iclrfinalcopy
\usepackage{hyperref}
\usepackage{url}
\usepackage{graphicx}

\title{Why So Inflammatory? Explainability in \newline Automatic Detection of Inflammatory Social Media Users}

\author{Cuong Nguyen, Daniel Nkemelu, Ankit Mehta \& Michael Best
\thanks{Presented at the PML4DC 2022 Workshop} \\
Georgia Institute of Technology\\
Atlanta, GA 30318, USA \\
\texttt{\{johnny.nguyen, dnkemelu, amehta318, mikeb\}@gatech.edu} \\
}

\begin{document}

\maketitle

\begin{abstract}
Hate speech and misinformation, spread over social networking services (SNS) such as Facebook and Twitter, have inflamed ethnic and political violence in countries across the globe. We argue that there is limited research on this problem within the context of the Global South and present an approach for tackling them. Prior works have shown how machine learning models built with user-level interaction features can effectively identify users who spread inflammatory content. While this technique is beneficial in low-resource language settings where linguistic resources such as ground truth data and processing capabilities are lacking, it is still unclear how these interaction features contribute to model performance. In this work, we investigate and show significant differences in interaction features between users who spread inflammatory content and others who do not, applying explainability tools to understand our trained model. We find that features with higher interaction significance (such as account age and activity count) show higher explanatory power than features with lower interaction significance (such as name length and if the user has a location on their bio). Our work extends research directions that aim to understand the nature of inflammatory content in low-resource, high-risk contexts as the growth of social media use in the Global South outstrips moderation efforts.

\end{abstract}

\section{Introduction}
Hate speech and disinformation spread on social networking services (SNS) continue to pose significant threats to the safety of people everywhere \citep{Unesco21}. Within the Global South, the spread of these types of problematic content has inflamed ethnic and political conflicts in several countries (the Rohingya genocide in Myanmar \citep{Fink18}, anti-Muslim riots in Sri Lanka and India \citep{Liebowitz21}, and anti-Hindu riots in Bangladesh \citep{Hasan21}). For this work, we refer to hate speech and misinformation used in these contexts as \textit{inflammatory content} due to their ability to inflame existing tensions into real-life violence, as highlighted in the cases above.

Tackling these issues in low-resource language settings presents significant challenges due to the unavailability of linguistic resources to build effective text classifiers. Given that the signals extracted from posts and tweets are invariably linked to an entity's profile, we hypothesize that focusing on and leveraging user-level features (such as in \citep{Ribeiro18, Elsherief18}) can help us understand and mitigate the spread of inflammatory content. These language-agnostic features bypass gaps in the availability and maturity of language-specific tools and resources.

In this paper, we explore an approach to detect users spreading inflammatory content using their interaction features and provide a detailed feature analysis of our classification model. We mainly focus on analyzing the explainability of classification models trained using these user-level features. Understanding model performance is required before deploying models that support tasks to moderate social media content. This engenders trust and accountability among platform owners, moderators and users. Our case study looks at Ethiopian Twitter users and the spread of inflammatory content during its contemporary civil and ethnic conflict. As all of Ethiopia's official languages, including Amharic, Oromo, and Tigrinya, are considered low-resource languages \citep{Hu20}, the context provides a prime example for situations where more conventional content-rich methods of analysis might fail thus motivating the methods we adopt in this work. 

\section{Methods}
\subsection{Data Collection}
This project was conducted in collaboration with the Center for Advancement of Rights and Democracy (CARD), an Ethiopian-based non-profit \citep{ICNL21}. To capture social media content related to the current Ethiopian civil conflict, we first identified a set of search terms (keywords and hashtags) associated with the conflict. We selected search terms that reflect the major ethnic and political groups involved in the ongoing conflict. We then used Twitter API's Filtered Stream endpoint, which pulls approximately one percent of publicly available tweets and their metadata, to collect tweets that contained any of our search terms. We collected tweets from August 17th, 2020, until July 29th, 2021. 

CARD trackers labelled a sample of the collected Twitter posts using the Aggie platform\footnote{https://aggie.readthedocs.io/en/latest/index.html}~\citep{Smyth16}. For consistency and contextual relevance, we adopted the definitions for hate speech and misinformation in the Ethiopian Government's 2020 Hate Speech Proclamation \citep{EthGov20}. We additionally referenced definitions of hate speech and misinformation provided by Twitter \citep{Ribeiro18}. At the end of the data collection process, we had collected 154 instances of inflammatory content that the trackers flagged. Out of those 154 instances, 58 posts (38 \(\%\)) had 'misinformation,' and 94 posts had 'hate speech' (61 \(\%\)) as specific type of inflammatory content. The remaining two posts (1 \(\%\)) either did not have a reason behind why it is inflammatory, or the reason did not fall into one of the two categories. We excluded these posts from further analysis. 

\subsection{Dataset Construction}
From the inflammatory posts found by the CARD trackers, we identified a set of inflammatory users (IUs). We defined IUs as any user who tweets or retweets one or more inflammatory posts, per the definition adopted above. We further subdivided these IUs into three subtypes based on the type of inflammatory content they spread. Hate Users (HUs) are users who tweet or retweet one or more posts flagged by trackers as hate speech. Misinfo Users (MUs) are users who tweet or retweet one or more posts flagged by trackers as misinformation, and Hate+Misinfo Users (HMUs) are users who are both Hate and Misinfo users. This strategy resulted in 589 IUs (145 HUs, 419 MUs, and 25 HMUs). 

To augment the set of IUs, we included users who used three or more instances of offensive and inflammatory terms from the PeaceTech Lab lexicon\citep{BarrachYousef21}, which we found to be highly correlated with the action of spreading inflammatory content. The lexicon was collected via surveys distributed to PeaceTech Lab's civil partners in Ethiopia and evaluated through online focus groups and expert interviews. We used the main portion of the lexicon for this particular study containing 21 terms and their linguistics variants. In the end, this resulted in a final set of 865 IUs (415 HUs, 409 MUs, 41 HMUs) 

Now that we have a set of inflammatory users (IUs), we constructed a set of non-inflammatory users (NIUs). This will enable us test differences between both groups based on our hypothesized user-level interaction features. This NIU sample captured users within our collected dataset that were active on Twitter during the time period but were not socially linked to the IUs previously identified. We assigned them to the non-inflammatory group based on homophily \citep{Ribeiro18, Mathew19}. We only selected users with ten or more unique activities (defined as either a tweet, a retweet, or a quoted tweet), which has been identified as a threshold for activeness in previous papers \citep{Ribeiro18, Mathew19}. This resulted in a smaller subset of 34,227 users. Then, we performed a diffusion process based on Degroot's Learning Model of the retweet graph similarly constructed from these users to \citep{Ribeiro18}. In the end, we selected 18,978 users with a belief score of 0 (i.e., did not retweet known IUs or people who retweet known IUs) to be our set of NIUs. The remaining 14,571 users with a non-zero belief score but do not belong to the IU group will be our set of borderline users (BUs).

\subsection{Characterization And Analysis Of Inflammatory Users}
We aim to identify and quantify the differences in profile metadata, usage patterns, and network centrality between NIUs, BUs, and IUs. To do this, we constructed a set of features that have been utilized in literature to identify differences between groups of Twitter users. Using the user feature classification framework presented in \citep{Volkova17}, we divide these features into 3 categories: profile features (e.g \textit{following} and \textit{follower}), syntactic and stylistic features (such as \textit{avg\_mention}, \textit{lex\_diversity}), network features (e.g \textit{eigencentrality}).

We used the Kruskal-Wallis test, alongside Dunn's posthoc test, to pinpoint where and how pairs of groups are significantly different for each feature. In addition to statistical significance, we are also interested in measuring the practical significance of any pairwise differences as they would inform how useful these features would be for building machine learning classifiers. To do so, we calculated Cliff's \(\delta\) for each feature and group pairs. We report the results of these statistical tests in table 1.

\subsection{Classification of Inflammatory Users}
Building on our hypothesis of significant differences between IUs and NIUs, we examine the feasibility of classifying users belonging to these two groups with traditional ML models and learning over graphs models that exploit Twitter's retweet network. For traditional ML models, we use Logistic Regression (LR), Support Vector Machine (SVM), Random Forest (RF), CatBoost (CB), and XGBoost (XGB) models. For learning over graphs models, we use GraphSAGE (GS) due to its proven effectiveness in node classification tasks \citep{Hamilton17}. For each user, we constructed a feature array of dimension 71 from these two categories of features: User features such as the number of followers, the number of activities, description length (21), and topic features, where each feature represents the prevalence of a given topic within the user's tweets (50). For GraphSage, we converted this user-feature matrix into a directed graph, with each node having a 71-dimensional vector representing the user. 

For both traditional ML and learning-over-graph methods, we randomly selected 80 percent of the data to be the training data and reserved the remaining 20 percent to be used as testing data. The randomized training and testing data were stratified so that for both sets, the ratio between IUs and NIUs was roughly identical to one another. We performed the classification process over ten random train-test stratified splits with five-fold cross-validation to tune the parameters of the classifiers. We use accuracy, macro-weighted precision, recall, and F1-score as the evaluation measures. To determine how the features contributed to the NIU/IU classification, we calculate their average SHAP (SHapley Additive exPlanations) values (on a logit scale) across all test splits and visualize these values with a beeswarm plot. SHAP uses a game-theoretic approach to explain the output of a machine learning model \citep{Lundberg2017}.
\newline\indent

\section{Results}
\subsection{Statistical Results}
We now present a synthesis of our statistical analysis results. See complete statistical results in table 1\footnote{p-values are from Dunn's post-hoc tests}. 

\subsubsection{Inflammatory Users (IUs) Are Influential}
The median IU has significantly higher \textit{following} and \textit{follower} compared to the median borderline user (BU) \((z = -36.83, p < 0.001)\), and even more so comparing to the median non-inflammatory user (NIU) \((z = -19.33, p < 0.001)\). At the median, IUs has three times the number of followings and fourteen times the number of followers compared to NIUs. The size of this disparity is further backed up by the Cliff's \(\delta\) for the NIU-IU pair from table 1, which indicates medium effect size for \textit{following} (\(\delta\) = -0.392) and large effect size for \textit{follower} (\(\delta\) = -0.57). In addition, we see that IUs are more influential within the retweet network compared to NIUs. They have significantly higher \textit{eigencentrality} than the other groups, scoring twice as much as NIUs (z = -12.74, p \(< 0.001\) \(\delta\) = -0.275) on median.

\subsubsection{Inflammatory Users Are Consistently Active}
We investigated differences in activity patterns between the groups. Not only did we calculate the raw number of activities per user, but we also calculated the ratio of their peak daily activity over their total number of unique activities or \textit{maxdate\(\_\)ratio}. We propose this feature as a simple proxy to users' continual engagement with topics relating to the current Ethiopian civil conflict. Users with a high \textit{maxdate\(\_\)ratio} concentrate most of their activities regarding the conflict on a single date. They are thus indicative of either fleeting engagement with the topic or bot-like activity (i.e., spamming tweets/retweets over a short period). As shown in table 1, we found that IUs have significantly higher \textit{activity\_count} and lower maxdate\(\_\)ratio (z = 21.15, p \(< 0.001\), \(\delta\) = 0.43), with five times as many unique activities and half the maxdate\(\_\)ratio compared to a NIU on median.

\begin{table}[]
\centering
\small
\begin{tabular}{|c|ccc|ccc|ccc|}
\hline
\textbf{Feature Name}   & \multicolumn{3}{c|}{\textbf{NIU-BU}}                                                                & \multicolumn{3}{c|}{\textbf{NIU-IU}}                                                                & \multicolumn{3}{c|}{\textbf{BU-IU}}                                                                   \\ \hline
\multicolumn{1}{|l|}{}  & \multicolumn{1}{c|}{z-score} & \multicolumn{1}{c|}{p} & Cliff’s \(\delta\)                 & \multicolumn{1}{c|}{z-score} & \multicolumn{1}{c|}{p} & Cliff’s \(\delta\)                  & \multicolumn{1}{c|}{z-score} & \multicolumn{1}{c|}{P-value} & Cliff’s \(\delta\)                   \\ \hline
following                  & \multicolumn{1}{c|}{-36.83}      & \multicolumn{1}{c|}{\(< 0.001\)}     & \cellcolor[HTML]{FFFF00}-0.23 & \multicolumn{1}{c|}{-19.33}      & \multicolumn{1}{c|}{\(< 0.001\)}     & \cellcolor[HTML]{FFA500}-0.39 & \multicolumn{1}{c|}{-7.61}       & \multicolumn{1}{c|}{\(< 0.001\)}     & \cellcolor[HTML]{FFFF00}-0.15   \\ \hline
eigencentrality         & \multicolumn{1}{c|}{7.45}        & \multicolumn{1}{c|}{\(< 0.001\)}     & 0.048                          & \multicolumn{1}{c|}{-12.74}      & \multicolumn{1}{c|}{\(< 0.001\)}     & \cellcolor[HTML]{FFFF00}-0.27 & \multicolumn{1}{c|}{-15.00}      & \multicolumn{1}{c|}{\(< 0.001\)}     & \cellcolor[HTML]{FFFF00}-0.28   \\ \hline
follower                & \multicolumn{1}{c|}{-52.05}      & \multicolumn{1}{c|}{\(< 0.001\)}     & \cellcolor[HTML]{FFA500}-0.33  & \multicolumn{1}{c|}{-27.75}      & \multicolumn{1}{c|}{\(< 0.001\)}     & \cellcolor[HTML]{FF0000}-0.57  & \multicolumn{1}{c|}{-11.19}      & \multicolumn{1}{c|}{\(< 0.001\)}     & \cellcolor[HTML]{FFFF00}-0.21   \\ \hline
description\_length     & \multicolumn{1}{c|}{-39.38}      & \multicolumn{1}{c|}{\(< 0.001\)}     & \cellcolor[HTML]{FFFF00}-0.24  & \multicolumn{1}{c|}{-15.21}      & \multicolumn{1}{c|}{\(< 0.001\)}     & \cellcolor[HTML]{FFA500}-0.29 & \multicolumn{1}{c|}{-2.72}       & \multicolumn{1}{c|}{0.01}     & -0.053                           \\ \hline
account\_age            & \multicolumn{1}{c|}{-27.79}      & \multicolumn{1}{c|}{\(< 0.001\)}     & \cellcolor[HTML]{FFFF00}-0.18 & \multicolumn{1}{c|}{-16.64}      & \multicolumn{1}{c|}{\(< 0.001\)}     & \cellcolor[HTML]{FFA500}-0.33 & \multicolumn{1}{c|}{-7.78}      & \multicolumn{1}{c|}{\(< 0.001\)}     & \cellcolor[HTML]{FFFF00}-0.16   \\ \hline
maxdate\_ratio          & \multicolumn{1}{c|}{30.01}       & \multicolumn{1}{c|}{\(< 0.001\)}     & \cellcolor[HTML]{FFFF00}0.19   & \multicolumn{1}{c|}{21.15}       & \multicolumn{1}{c|}{\(< 0.001\)}     & \cellcolor[HTML]{FF0000}0.43   & \multicolumn{1}{c|}{11.56}       & \multicolumn{1}{c|}{\(< 0.001\)}     & \cellcolor[HTML]{FFFF00}0.23    \\ \hline
avg\_mention            & \multicolumn{1}{c|}{3.93}        & \multicolumn{1}{c|}{\iffalse0.0001\fi \(< 0.001\)}     & 0.024                        & \multicolumn{1}{c|}{11.39}       & \multicolumn{1}{c|}{\(< 0.001\)}     & \cellcolor[HTML]{FFFF00}0.239  & \multicolumn{1}{c|}{10.08}       & \multicolumn{1}{c|}{\(< 0.001\)}     & \cellcolor[HTML]{FFFF00}0.19     \\ \hline
avg\_hashtag            & \multicolumn{1}{c|}{0.45}          & \multicolumn{1}{c|}{0.98}     & 0.0025                         & \multicolumn{1}{c|}{8.93}       & \multicolumn{1}{c|}{\(< 0.001\)}     & \cellcolor[HTML]{FFFF00}0.18  & \multicolumn{1}{c|}{8.73}       & \multicolumn{1}{c|}{\(< 0.001\)}     & \cellcolor[HTML]{FFFF00}0.17    \\ \hline
follower\_following\_ratio & \multicolumn{1}{c|}{-43.83}      & \multicolumn{1}{c|}{\(< 0.001\)}     & \cellcolor[HTML]{FFFF00}-0.28 & \multicolumn{1}{c|}{-24.51}      & \multicolumn{1}{c|}{\(< 0.001\)}     & \cellcolor[HTML]{FF0000}-0.50 & \multicolumn{1}{c|}{-10.56}      & \multicolumn{1}{c|}{\(< 0.001\)}     & \cellcolor[HTML]{FFFF00}-0.20   \\ \hline
name\_length            & \multicolumn{1}{c|}{10.84}       & \multicolumn{1}{c|}{\(< 0.001\)}     & 0.069                          & \multicolumn{1}{c|}{3.87}       & \multicolumn{1}{c|}{\iffalse0.0002\fi \(< 0.001\)}     & 0.077                         & \multicolumn{1}{c|}{0.44}           & \multicolumn{1}{c|}{0.99}     & 0.01                             \\ \hline
avg\_tweet\_length      & \multicolumn{1}{c|}{32.21}       & \multicolumn{1}{c|}{\(< 0.001\)}     & \cellcolor[HTML]{FFFF00}0.20  & \multicolumn{1}{c|}{15.65}       & \multicolumn{1}{c|}{\(< 0.001\)}     & \cellcolor[HTML]{FFA500}0.31  & \multicolumn{1}{c|}{5.40}        & \multicolumn{1}{c|}{\(< 0.001\)}     & \cellcolor[HTML]{FFFF00}0.11    \\ \hline
avg\_url                & \multicolumn{1}{c|}{10.28}       & \multicolumn{1}{c|}{\(< 0.001\)}     & 0.065                          & \multicolumn{1}{c|}{-5.14}       & \multicolumn{1}{c|}{\(< 0.001\)}     & -0.10                         & \multicolumn{1}{c|}{-8.34}       & \multicolumn{1}{c|}{\(< 0.001\)}     & \cellcolor[HTML]{FFFF00}-0.17   \\ \hline
Bot\_score              & \multicolumn{1}{c|}{20.71}       & \multicolumn{1}{c|}{\(< 0.001\)}     & \cellcolor[HTML]{FFFF00}0.13   & \multicolumn{1}{c|}{2.72}        & \multicolumn{1}{c|}{0.001}     & 0.061                          & \multicolumn{1}{c|}{-3.82}       & \multicolumn{1}{c|}{\iffalse0.0002\fi \(< 0.001\)}     & -0.086                           \\ \hline
activity\_count       & \multicolumn{1}{c|}{-18.67}       & \multicolumn{1}{c|}{\(< 0.001\)}     & -0.12                          & \multicolumn{1}{c|}{-20.39}      & \multicolumn{1}{c|}{\(< 0.001\)}     & \cellcolor[HTML]{FFA500}-0.398 & \multicolumn{1}{c|}{-14.38}      & \multicolumn{1}{c|}{\(< 0.001\)}     & \cellcolor[HTML]{FFA500}-0.305 \\ \hline
lex\_diversity          & \multicolumn{1}{c|}{3.43}       & \multicolumn{1}{c|}{\(< 0.001\)}     & 0.022                          & \multicolumn{1}{c|}{-13.22}      & \multicolumn{1}{c|}{\(< 0.001\)}     & \cellcolor[HTML]{FFFF00}-0.27 & \multicolumn{1}{c|}{-14.31}      & \multicolumn{1}{c|}{\(< 0.001\)}     & \cellcolor[HTML]{FFA500}-0.28 \\ \hline
\end{tabular}
\caption{\bf{z-score, p-value and effect size (using Cliff's \(\delta\)) from the Dunn's post-hoc test (with Bonferroni correction) between pairs of groups. Here, yellow cells represents small effect size, orange represents medium effect size, and red cells represents large effect size in accordance to the reference scale mentioned above}}
\end{table}
\subsection{Classification Results}
We used the trained model to predict whether users in the test set belonged to the IU or the NIU group. The results were aggregated from the 10 train-test splits, and the means of aforementioned evaluation measures are reported in Table 2.  
\newline
\begin{table}[]
\centering
\small
\begin{tabular}{|c|c|c|c|c|}
\hline
Method    & Precision      & Recall         & F1             & Accuracy       \\ \cline{0-4}
Support Vector Machine (SVM)       & 0.478 & 0.500& 0.489 & 0.956\\
Random Forest (RF)        & 0.930 & 0.753 & 0.816 & 0.975 \\
Logistic Regression (LR)       & 0.804 & 0.579 & 0.619 & 0.959 \\
XGBoost (XGB)  & 0.929 & \textbf{0.767} & \textbf{0.827}  & \textbf{0.976} \\
CatBoost (CB) & \textbf{0.957}  & 0.705 & 0.779 & 0.973 \\
GraphSage (GS) & 0.585              & 0.546              & 0.565              & 0.96             \\ \hline
\end{tabular}
\caption{\textbf{Evaluation metrics for the task of predicting if a user is a IU or a NIU. The values reported are the means over the 10 random train-test splits for each of the evaluation metric (macro-weighted)}} 
\end{table}
We got the best performance from the XGBoost model using topic and user features, achieving an F1 of 0.827. Classifiers such as Random Forest and CatBoost also performed comparably well. The standard deviation for the evaluation metrics ranges from 0.001 to 0.02, from which we conclude that our results are relatively consistent across train-test splits. Learning-over-graph methods such as GraphSage did not perform as well as traditional ML methods (0.565 vs. 0.827 for XGBoost). We attribute this to the class imbalance (1:20) between IUs and NIUs, which seemed to affect learning-over-graph methods compared to traditional ML methods. We also notice that the macro-averaged precision is higher across the board than macro-averaged recall, signaling that our models were accurate yet conservative when assigning users into the IU group. Furthermore, an analysis of the beeswarm plot in Figure 1 reveals that features with high practical significance (such as activity\_count, follower and maxdate\_ratio) also have high SHAP values in the appropriate direction. In contrast, features with low practical significance (such as has\_location, name\_length) have SHAP values closely bunched around 0 (i.e little impact on model output).  
\begin{figure}
    \includegraphics[scale=0.45]{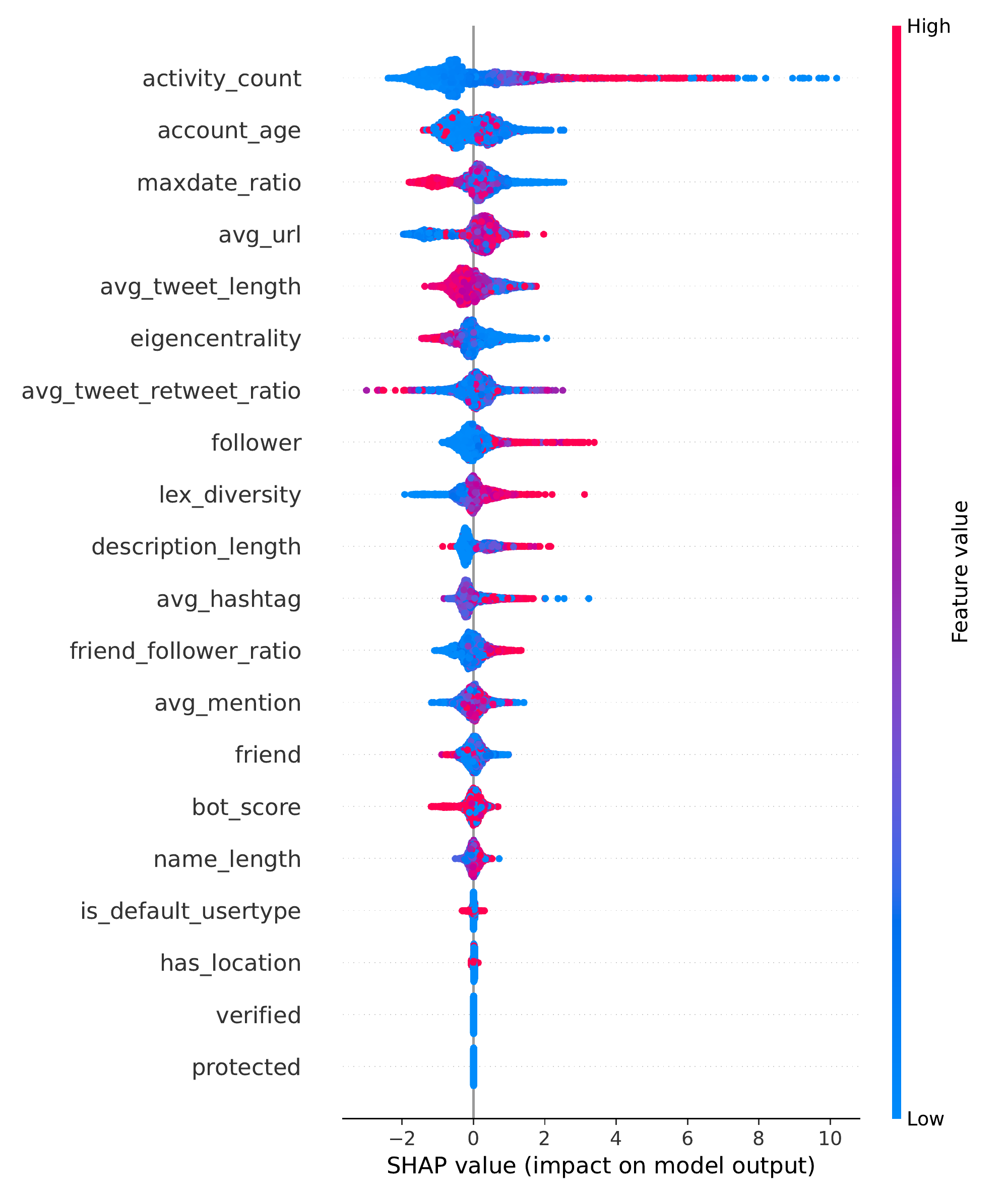}
    \caption{\textbf{Beehive plot representing SHAP values for user-level features generated on the test-set by the XGBoost model. Each dot represents an user with the test set. The x-axis represents the SHAP value the model assigned to that feature given the user classification. Positive values indicate increased likelihood of IU prediction and negative values indicate increased likelihood of NIU prediction (on a logit scale)}}
    \label{fig:my_label}
\end{figure}

\subsection{Conclusion}
This paper presents an approach to characterize and detect inflammatory content on Twitter using user-level interaction metrics. We argue that this method is particularly effective for contexts in the Global South where natural language ground truth data and processing resources are scarce. We find that inflammatory users - users that have been identified to share hate speech and disinformation posts - show significant distinctions from non-inflammatory users on key interaction metrics. They are more influential, active, and non-botlike in their interaction. They also tend to post more diverse content within the discussion surrounding the conflict. Extending our statistical analyses of group differences with SHAP analysis of our XGBoost model, we found that the best model trained to distinguish IUs from NIUs utilized features with high practical significance from the statistical analysis. This work extends research directions that aim to understand the nature of inflammatory content in low-resource contexts as the growth of social media use in the Global South overwhelms platform moderation.

\bibliography{iclr2022_conference}
\bibliographystyle{iclr2022_conference}

\end{document}